\documentclass[aps,twocolumn,groupedaddress,floatfix]{revtex4-1}
\usepackage{amssymb,amsmath}
\usepackage{graphicx}
\usepackage{subfigure}
\usepackage[english]{babel}
\usepackage{float}
\usepackage{color}

\usepackage{lipsum}
\begin{document}
\newcommand{\be}{\begin{equation}}
\newcommand{\ee}{\end{equation}}
\newcommand{\rojo}[1]{\textcolor{red}{#1}}

\title{Discrete embedded modes in the continuum in 2D lattices}

\author{Mario I. Molina}
\affiliation{Departamento de F\'{\i}sica, Facultad de Ciencias, Universidad de Chile, Casilla 653, Santiago, Chile}

\date{\today }

\begin{abstract} 
We study the problem of constructing bulk and surface embedded modes (EMs) inside the quasi-continuum band of a square lattice, using a potential engineering approach \`a la Wigner and von Neumann. Building on previous results for the one-dimensional (1D) lattice, and making use of separability, we produce examples  of two-dimensional envelope functions and the two-dimensional (2D) potentials that produce them.  The 2D embedded mode  decays like a stretched exponential, with a supporting potential that decays as a power law. The separability process can cause that a  1D impurity state (outside the 1D band) can give  rise to a 2D embedded mode (inside the band). The embedded mode survives the addition of random perturbations of the potential; however, this process introduces other localized modes inside the band, and causes a general tendency towards localization of the perturbed modes. 

\end{abstract}

\maketitle

\section{Introduction}
One of the most well-known results in quantum mechanics dictates that, for a particle in the presence of a potential, its state is either a localized one, with negative energy, forming a discrete spectrum, or an  extended mode, with a quasi-continuum spectrum, where  the energy is positive. This  result which stems from a basic analysis of the Scr\"{o}dinger equation, was challenged by Wigner and von Neumann in 1929\cite{wigner}. They  proposed that another case could exist: That of a localized mode whose energy falls inside the band continuum. This embedded mode ({\bf EM}) is decoupled from the extended ones and has infinite lifetime, like a resonance with zero width. Using a reverse engineering approach, they were capable of building a spherically symmetric 3D continuous potential capable of rendering a given free particle mode as localized, with a power law decay in space. Regarded at first as a mathematical curiosity, the subject has emerged recently with a substantial number of works and application in many diverse areas where wave phenomena are dominant.  In the seventies, Stillinger\cite{stillinger} and Herrick\cite{herrick} suggested that EMs might be found in certain atomic and molecular systems. They also suggested the use of superlattices to construct potentials that could support EMs\cite{superlattice1, superlattice2}. Direct observation of electronic bound states above a potential well localized due to Bragg reflections, were carried out using semiconductor heterostructures\cite{capasso}. A different approach to EMs comes from the Physics of resonant states in quantum mechanics. These are states that are localized in space but with non-decaying wings and energies inside the band and they eventually decay, i.e., they possess a finite lifetime. Sometimes, these resonances can interfere with each other giving rise to a resonance of zero width, that is, an EM. This can happen to a Hydrogen atom inside a magnetic field, modeled as a system of coupled Coulombic channels, where interference between resonances belonging to different channels can lead to the creation of an EM\cite{wintgen1,wintgen2}. In recent times, EMs have been shown to occur in mesoscopic electron transport and quantum waveguides\cite{electron1,electron2,electron3,electron4,electron5,electron6,electron7,electron8,electron9}, and in quantum dot systems\cite{orellana1,orellana2,orellana3,orellana4,orellana5}. Not surprisingly, here the existence of EMs can be traced back to the destruction of the discrete-continuum decay channels by quantum interference effects.

The origin of the EM phenomenology is regarded nowadays as the results of interference and thus it should be inherent to any wavelike theory besides quantum mechanics, such as classical optical systems described by the paraxial wave equation. Recent use of the analogy between these two realms have proven fruitful, leading to the observation of many phenomena that are hard to observe in a condensed matter setting, such as dynamic localization\cite{dynamic}, Bloch oscillations\cite{bloch1,bloch2},Zeno effect\cite{zeno1,zeno2} and Anderson localization\cite{anderson1,anderson2,anderson3,anderson4}, to name some. The appeal of using optical systems is that the experiment can be designed to focus on a particular aspect, bypassing the need to deal with other affects commonly present in condensed matter, such as many-body effects\cite{longhi}. The ability to steer optical excitations and tailor the optical medium, makes this type of system attractive for studies of EMs\cite{optics1,optics2,optics3,optics4,optics5,optics6}. A  comprehensive review on EMs and its properties and applications can be found in ref. \cite{marin}.

In this work we study the problem of constructing an embedded mode (EM) inside the band of a two-dimensional (2D) square lattice, using a potential engineering approach, following Wigner and von Neumann, extended to a discrete, periodic system. The general methodology consists on choosing an envelope that modulates a given chosen extended eigenstate, and imposing that its energy coincides with the original one (no envelope). To achieve this, we introduce a site energy distribution $\epsilon_{\bf n}$ chosen precisely for leading to an energy for the modulated state that coincides with the energy of the original mode. This intimate association between potential and mode is reminiscent of supersymmetry, where a Darboux transformation is effected on a free particle state (extended) to yield a different potential where the corresponding state is localized but retains its positive energy, i.e., remains in the continuum\cite{SUSY1,SUSY2}.

\section{Model}
Let us consider a generic excitation propagating in a D-dimensional tight-binding lattice, in the absence of a local potential. The stationary equations for the mode amplitudes have the form:
\be
-\lambda \ \phi_{\bf n} + \sum_{{\bf m}\neq{\bf n}} V_{{\bf n},{\bf m}} \phi_{\bf m} = 0\label{eq1}
\ee
where $\phi_{\bf n}$ is the mode amplitude at site ${\bf n}$, $V_{{\bf n},{\bf m}}$ is the coupling between sites ${\bf n}$ and ${\bf m}$, and $\lambda$ is the eigenvalue. An interesting physical realization of this system consists on a set of weakly-coupled optical waveguides, 
each of them centered at lattice site ${\bf n}$. In the coupled-modes approach, the electrical field $E(\vec{x},z)$ inside the guide is expanded as $E(\vec{x},z)=\sum_{\bf n} \phi_{\bf n}\ \psi(\vec{x}-{\bf n})$, where $\psi(\vec{x})$ is the waveguide mode and $z$ is the longitudinal propagation distance along the guide. After replacing into the paraxial wave equation and looking for stationary modes, $\phi_{\bf n}(z)= \phi_{\bf n}\ \exp(i \lambda z)$, we arrive at Eq.(\ref{eq1}). 

Next, we proceed to modulate a chosen state $\phi_{\bf n}$ as $C_{\bf n} = f_{\bf n} \phi_{\bf n}$ and impose the condition that this modulated state be an eigenstate of the system with the {\em same} eigenvalue. To do this, we need to introduce a distribution of propagation constants $\epsilon_{\bf n}$ in Eq.(\ref{eq1}):
\be
( -\lambda + \epsilon_{\bf n} ) \ f_{\bf n} \phi_{\bf n} + \sum_{{\bf m}\neq{\bf n}} V_{{\bf n},{\bf m}} f_{\bf m}\phi_{\bf m} = 0,\label{eq2}
\ee
from which it is possible to formally express
\be
\epsilon_{\bf n} = \lambda -\sum_{{\bf m}\neq{\bf n}} V_{{\bf n},{\bf m}} (f_{\bf m}\phi_{\bf m}/f_{\bf n}\phi_{\bf n}).
\label{eq3}
\ee
Thus, for a given envelope $f_{\bf n}$, it is possible to obtain the distribution of propagation constants $\epsilon_{\bf n}$ that allows the modulated state to remain as an eigenstate of the system with the same eigenvalue as the unmodulated state. The envelope function $f_{\bf n}$ decays away from a chosen site and it should be normalizable: $\sum_{\bf n}f_{\bf n}^2<\infty$.

We seemed to have found a simple way to convert an extended state into a localized one; however there are some important details that need to be addressed. For instance, we can see from Eq.(\ref{eq3}) that it is possible for $\epsilon_{\bf n}$ to diverge at the zeroes of $\phi_{\bf n}$. This is quite possible given that the modes of the unperturbed system oscillate in space. Since divergences or near-divergences in the site propagation constants distribution are undesirable, our choice of $f_{\bf n}$ must avoid them. This is a no trivial condition which makes us look in more detail at what happens near the zeroes of $\phi_{\bf n}$. This methodology has been applied in the past to build surface and bulk EMs for a discrete periodic one-dimensional tight-binding chain, that were structurably stable and whose EM eigenvalue was tunable by means of nonlinearity\cite{1D1}.

\section{Results}

For the particular case of a square lattice, Eq.(\ref{eq1}) can be cast as:
\be
-\lambda C_{n,m} + V\ ( C_{n+1,m}+C_{n-1,m}+C_{n,m+1}+C_{n,m-1} ) = 0\label{eq4}
\ee
Using the separability property of the square lattice, 
Eq.(\ref{eq3}) becomes completely separable and $C_{n,m} = C_{n}C_{m}$ and $\lambda=\lambda_{a} +\lambda_{b}$, and $\epsilon_{n,m} = \epsilon_{n} + \epsilon_{m}$,
where
\begin{eqnarray}
( -\lambda^{a} + \epsilon_{n} )\ C_{n}^{a} + V ( C_{n+1}^{a}+C_{n-1}^{a})&= &0\nonumber\\
( -\lambda^{b} + \epsilon_{m} )\ C_{m}^{b} + V ( C_{m+1}^{b}+C_{m-1}^{b})&= &0\label{eq5}
\end{eqnarray}
where,``a'' and ``b'' denote states of a one-dimensional chain. The problem is thus reduced to that of two uncoupled one-dimensional (1D) chains. The case of a chain  
\begin{figure}[t]
 \includegraphics[scale=0.215]{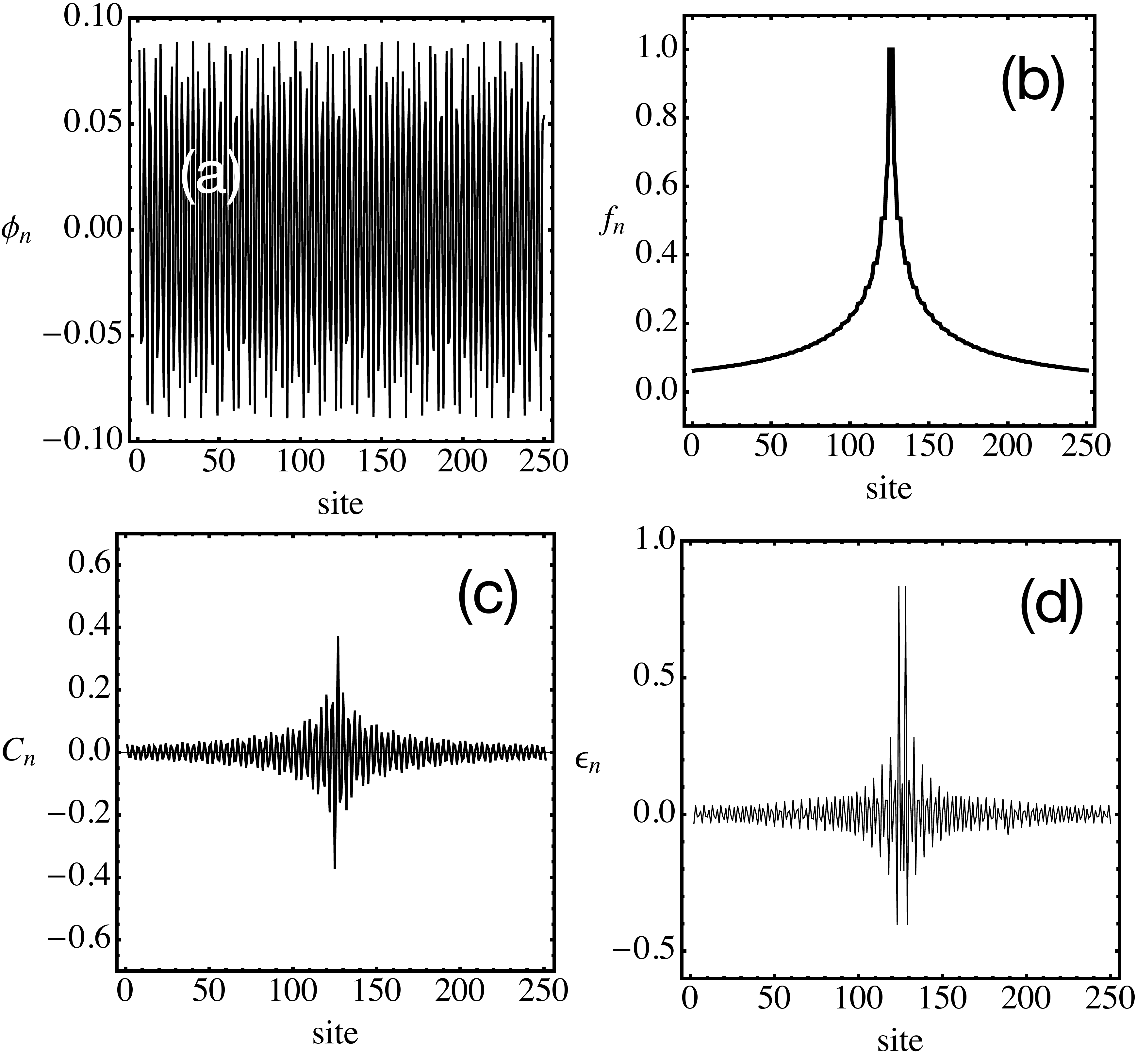}
 \caption{Generation of a 1D localized mode inside the band. (a) Unmodulated state (b) Modulating envelope (c) Modulated state (d) Local potential needed. ($N=251, n_{c}=125$) }
  \label{fig1}
\end{figure}
has been addressed before\cite{1D1, 1D2}. We reproduce its main features here, for the sake of completeness.

For a 1D chain of length $N$, Eqs.(\ref{eq3}) for the distribution of propagation constants become
\begin{eqnarray}
\epsilon_{1} &=& \lambda - V {f_{2}\over{f_{1}}} \ {\phi_{2}\over{\phi_{1}}},\nonumber\\
\epsilon_{n} &=& \lambda - V {f_{n+1}\over{f_{n}}} \ {\phi_{n+1}\over{\phi_{n}}}- V         {f_{n-1}\over{f_{n}}} \ {\phi_{n-1}\over{\phi_{n}}},\hspace{0.5cm}1<n<N\nonumber\\
\epsilon_{N} &=&  \lambda - V {f_{N-1}\over{f_{N}}} \ {\phi_{N-1}\over{\phi_{N}}}
\end{eqnarray}
Assuming  a decreasing envelope around a site $n_{0}$, we can write
\be
{f_{n+1}\over{f_{n}}} = 1-\delta_{n}
\ee
to the right of $n_{0}$, with $\delta_{n}<1$. To the left of $n_{0}$, we have
\be
{f_{n}\over{f_{n+1}}} = 1-\delta_{n}.
\ee

These expressions lead to
\be
f_{n} = \prod_{m=1}^{|n-nc|-1} (1-\delta_{m})
\ee
\begin{figure}[t]
 \includegraphics[scale=0.25]{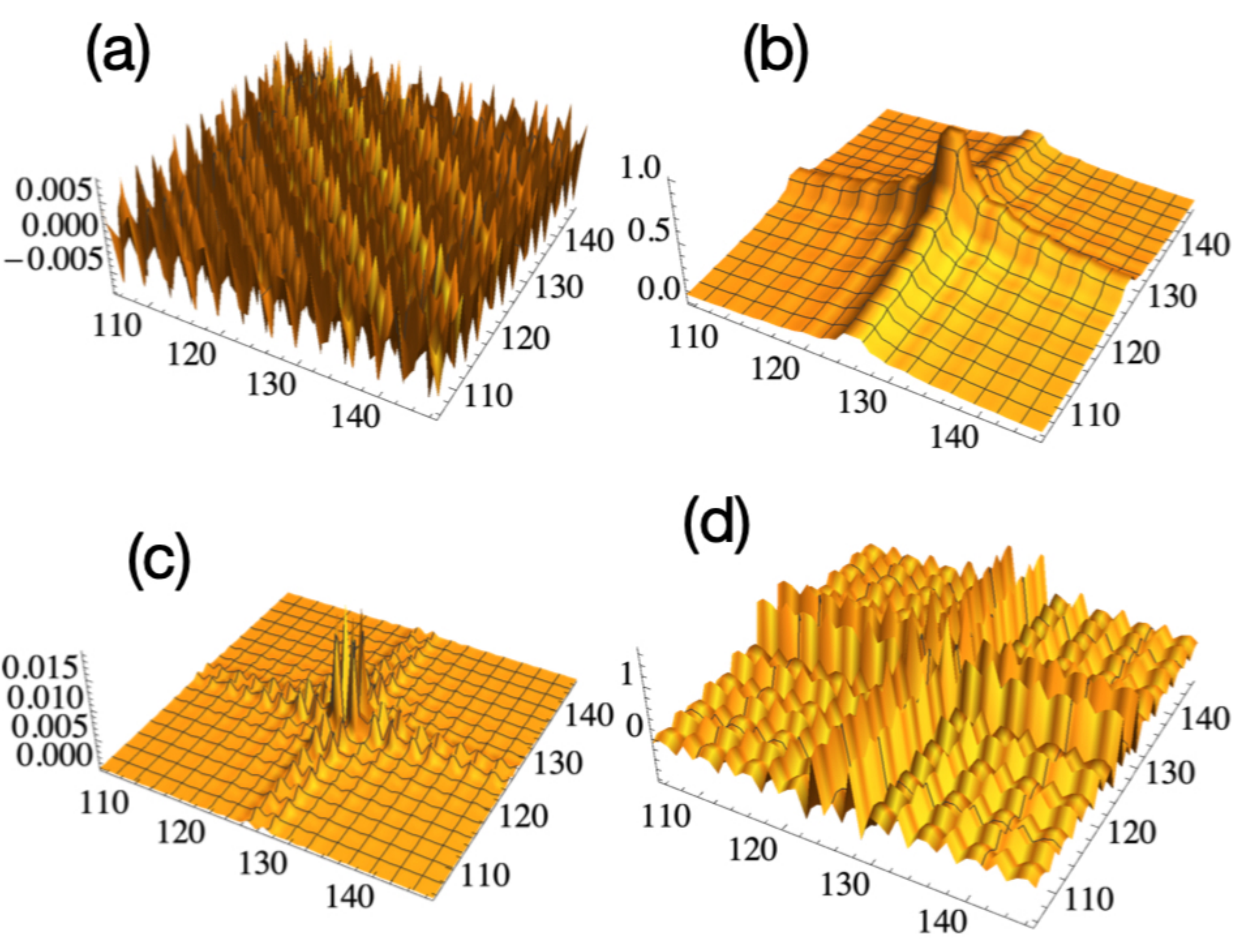}
 \caption{Generation of a 2D localized bulk mode inside the band. (a) Unmodulated state (b) Modulating envelope (c) Modulated state (d) Local potential needed. ($N=251\times 251, n_{c}=125$) }
  \label{fig2}
\end{figure}
\begin{figure}[h]
 \includegraphics[scale=0.22]{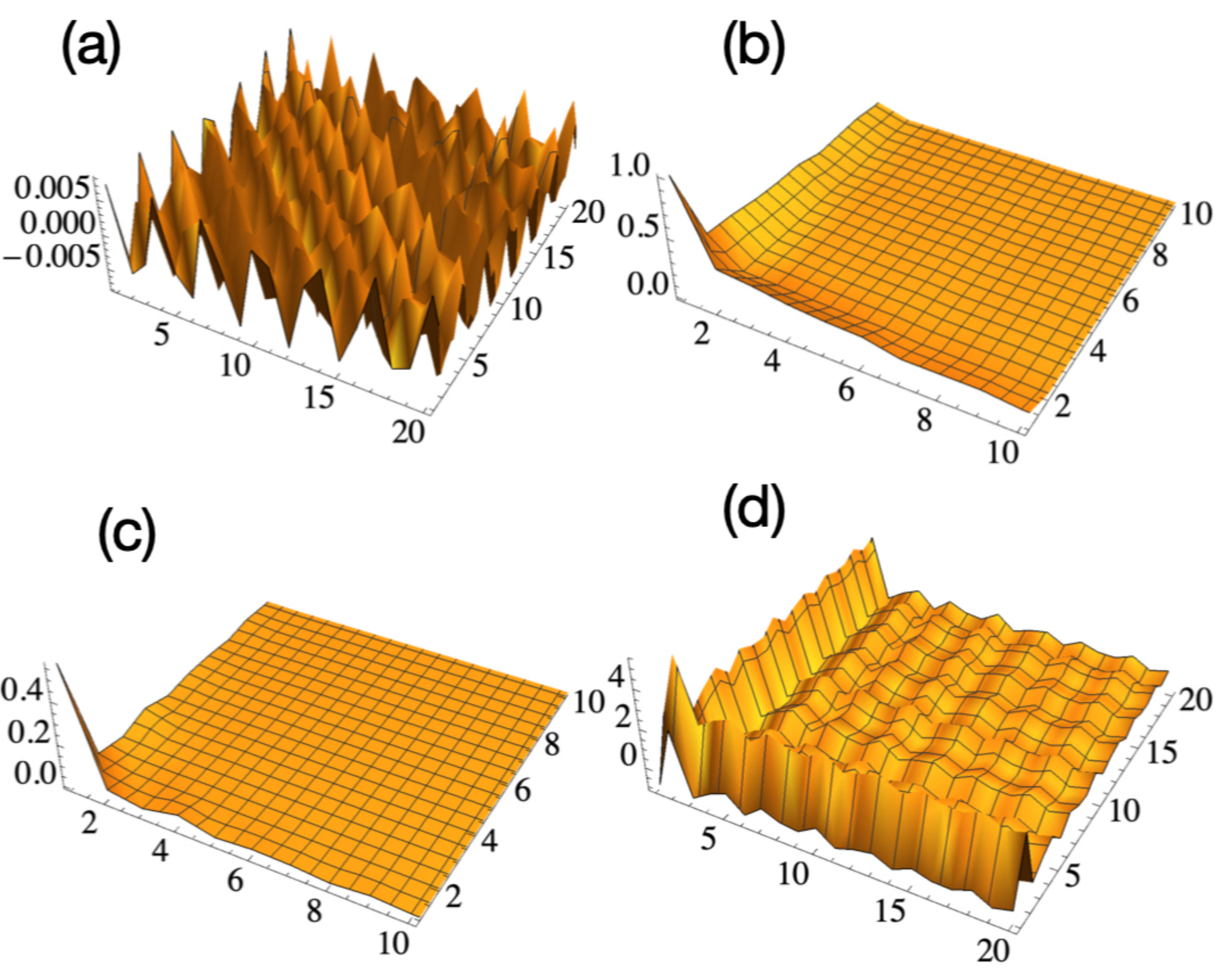}
 \caption{Same as in Fig.2 but for an EM placed at the corner of a $251\times 251$ square lattice. }
  \label{fig3}
\end{figure}

Now, how do we choose $\delta_{n}$? Ideally, we would like to have a $f_{n}$ that is bounded and smooth. We want to obtain a localized mode without disturbing the rest of the band modes too much, so they remain extended. This means, to have a distribution $\{\epsilon_{n}\}$ that is also ``smooth'' and bounded. Simple analysis\cite{1D1} shows that, in order to have a normalizable envelope, one needs $\sum_{n}\delta_{n}\rightarrow\infty$ with $\delta_{n}<1$. The other requirement is that $f_{n}$ does not diverge at the zeroes of $\phi_{n}$. A possible choice that fulfills these conditions is \cite{1D1}
\be
\delta_{n} = {a\over{1+|n-n_{c}|^{b}}}\ N^2 \phi_{n}^2 \phi_{n+1}^2,
\ee
\begin{figure}[t]
 \includegraphics[scale=0.22]{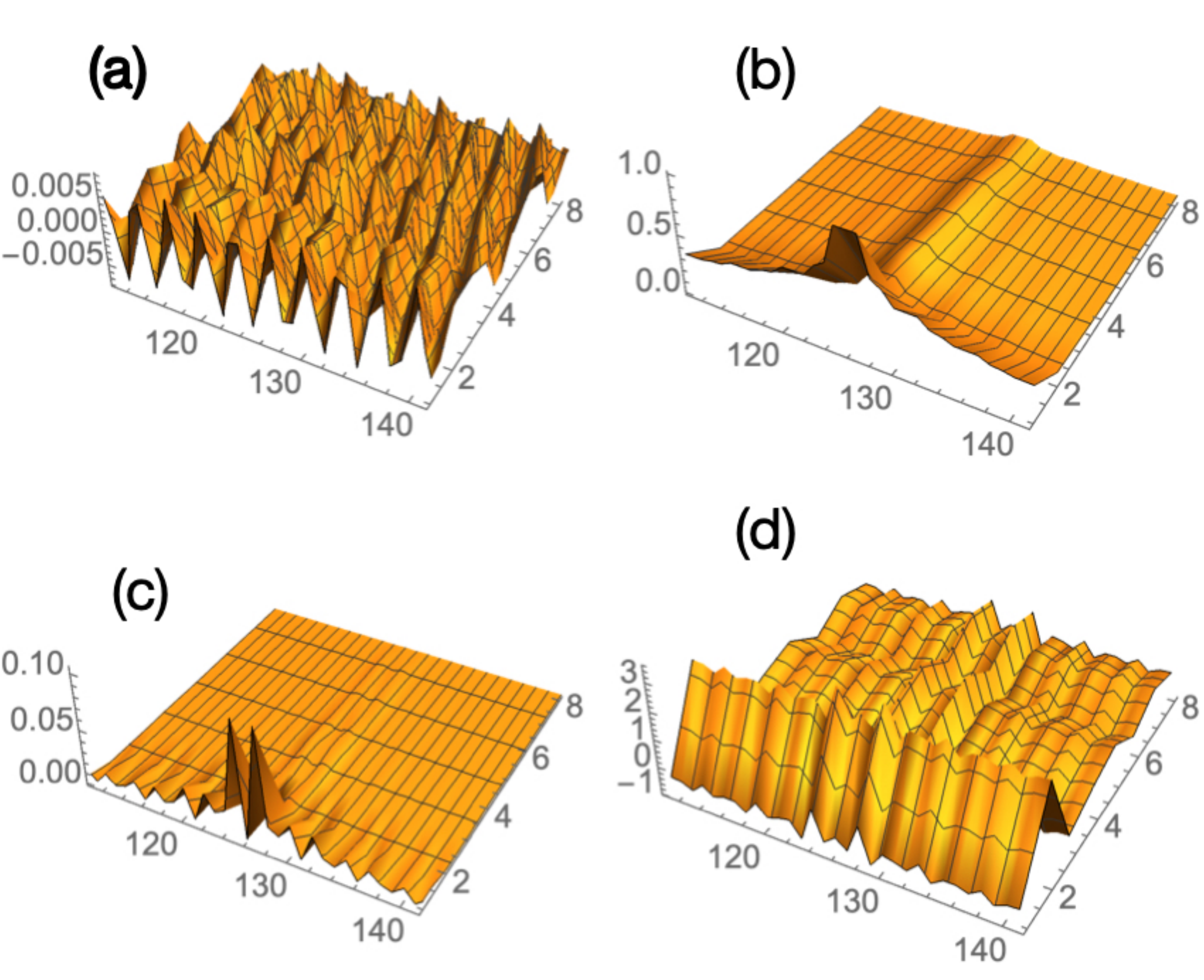}
 \caption{Same as in Fig.2 but for an EM placed at the edge of a $251\times251$ square lattice}
  \label{fig4}
\end{figure}
with $b<1$.  The presence of $N$ is due to the normalization of the $\phi_{n}\sim 1/\sqrt{N}$. This $\delta_{n}$ decays as a power law, and oscillates in space. The oscillating terms are needed to counteract the effect of the zeroes of $\phi_{n}$ in Eq.(\ref{eq3}), in such a way to ensure that near the zeroes of 
$\phi_{n}$, $f_{n}$ remains finite. In fact, with this choice, $\epsilon_{n}$  is identically zero at the zeroes of $\phi_{n}$.  Figure \ref{fig1} shows results for an example of a 1D chain, treated by this procedure. It was observed that, besides the EM, some impurity modes were created, as well as some resonant states, which are characterized for having a central maximum with wings that do not decay in space.
A hand waving argument as to how the EM is produced physically, could be as follows: Notice the bristly nature of $\epsilon_{n}$, with strong fluctuations in amplitude from site to site. Imagine now placing a localized excitation in the form of the EM  at $n_{c}$. It will try to expand away and thus will experience scattering from the potential. Roughly speaking, the potential has the form of a collection of impurity potentials, each of which will scatter the expanding mode. Because of the clever way the potential is built, different portions of the  wave will get scattered in such a way as to lead to an ever decreasing transmission coefficient, because of purely negative interference. The EM is a wave that tries to be extended but it cannot, due to interference effects. 

Let us go now to the two-dimensional square lattice.  Because of complete separability, we have $C_{n,m}=C_{n} C_{m} = f_{n} f_{m} \phi_{n} \phi_{m}$, and $\epsilon_{n,m}=\epsilon_{n}+\epsilon_{m}$, where $f_{n}$ and $\epsilon_{n}$ correspond to the envelope and potential function for the 1D case. Results of this procedure are shown in Fig.\ref{fig2}, for a bulk localized mode in the square lattice, that is, a localized mode centered around a site that is far from the boundaries. The chosen unmodulated state in this example is the product of two unmodulated ones, each one chosen as the same 1D state shown in Fig.\ref{fig1}a. Thus, the 2D EM is just the product of two 1D EMs along two perpendicular directions.
The asymptotic decay of the 2D envelope function at large distances from the center of the envelope, $(n_{c},m_{c})$, can be obtained with the help of the Euler-Mclaurin expansion, to yield
\be
C_{n, m} \sim \exp(-\alpha(k_{x})|n-n_{c}|^{1-b}-\alpha(k_{y}) |m-m_{c}|^{1-b})
\label{eq11}
\ee
where, $\alpha(k)=(1/8(1-b))(2+\cos(2 k))$, and we have used $\phi_{n}=(1/\sqrt{N}) \sin(k n)$. The decay is in the form of a stretched exponential. The site potential needed to create this EM can be computed from Eqs.(\ref{eq3}) and (\ref{eq11}), yielding the asymptotic form
\be
\epsilon_{n,m}\approx {\gamma(k_{x})(1-b)\over{|n-n_{c}|^b}}+{\gamma(k_{y})(1-b)\over{|m-m_{c}|^b}}
\ee
with $\gamma(k)=2 \alpha(k) V \sin(k)\cot(n k)$ and $0<b<1$
i.e., a power law decay.
 
Figure 3 shows results for a surface embedded mode centered at one of the corners of the square lattice. In this case, we can choose 
\be
\delta_{n}=(a/\sqrt{n}) N^2 \phi_{n}^2 \phi_{n+1}^2,\hspace{0.5cm}1\leq n<N-1,
\ee
which implies 
\be
f_{n} = \left\{ \begin{array}{ll}
		\prod_{m=1}^{n-1} \ (1 - \delta_{m}) & \mbox{for $n>1$}\\
		\\
		1 &\mbox{for $n=1$,}
		\end{array}
		\right.		
\ee
leading to a two-dimensional embedded mode that decays as  
\be 
C_{n, m} \sim \exp(-\beta(k_{x}) \sqrt{n}-\beta(k_{y})\sqrt{m} ), \hspace{0.2cm} 0<n,m<N
\ee
with $\beta(k) = (1/4)(2 + \cos(2 k))$. Again a stretched exponential form. This mode is the product of two `surface' EMs along two perpendicular directions. The asymptotic decay of the potential has the form
\be
\epsilon_{n,m}\approx (1/2){\gamma(k_{x})\over{\sqrt{n}}}+(1/2){\gamma(k_{y})\over{\sqrt{m}}}.
\ee

Finally, for the case when the center of the EM is placed at one of the edges of the square lattice (Fig.\ref{fig4}), the EM can be regarded as a product of an 1D `bulk' EM centered at $n=n_{c}$ and another 1D `surface' EM that decays from $n=1$. Its asymptotic decay has the form
\begin{eqnarray}
C_{n, m} &\sim & \exp(-\alpha(k_{x})|n-n_{c}|^{1-b}-\beta(k_{y}) \sqrt{m}\ \ ),\\
         &     & \hspace{2.0cm} -N<n<N, 0<m<N
\end{eqnarray}
with a potential that decays at long distances as
\be
\epsilon_{n,m}\approx {\gamma(k_{x})(1-b)\over{|n-n_{c}|^b}}+(1/2){\gamma(k_{y})\over{\sqrt{m}}}.
\ee

In all cases, the decaying envelope has the form of a stretched exponential. This is faster than in Wigner and von Neumann's power law case, but slower than a pure exponential. The potential decays as a power law, in all cases.

It should be emphasized that the detailed shape of $\epsilon_{n,m}$ depends strongly on $f_{n,m}$. This means that if we change the envelope parameters, the local potential could become less ``ideal'' (bristlier or with stronger fluctuations) than the cases shown in Fig.\ref{fig2},\ref{fig3} and \ref{fig4}. It would still be a {\em bona fide} solution, just less smooth. Thus, there is plenty of room to play with the system parameters, to arrive at the desired mode.

At this point, one could but wonder about the effect of this nontrivial $\epsilon_{n,m}$ on the rest of the states of the band. As we said before, all depends on the details of the envelope function $f_{n,m}$. If it is smooth on the scale of the lattice, it will give rise to a smooth self-consistent potential $\epsilon_{n,m}$. This means that the rest of the modes in the band will be only slightly perturbed, and will retain their character as extended states (no longer sinusoidal, though). On the contrary, if the envelope changes quickly on the scale of the lattice, it will give rise to a potential that is quite bristlier with strong height oscillations. Under these conditions, the potential looks like a sum of impurity potentials and thus, gives rise to a number of impurity-like states which are localized, but fall outside the band. In the case when one neglects the oscillatory component of $f_{n,m}$ and retains only the envelope, there will be strong fluctuations and divergencies or near divergences in $\epsilon_{n,m}$ due to the zeroes of $\phi_{n,m}$ (See Eq.(\ref{eq3})). This complicated form for $\epsilon_{n,m}$ might mimic a random potential and thus   cause the states of the band (besides the EM) to become localized.

All of the above discussion hinges around the concept of localization: How to create a localized state and leave the rest extended, how a poor choice of potential will bring about many more localized modes, how some resonant states might be created, or how --in a limit case--all of the states will be localized \`{a} la Anderson. A common measure of the degree of localization of a state is its participation ratio $R$,
\be
R(t) = {( \sum_{\bf n} |C_{\bf n}(t)|^2  )^2\over{\sum_{\bf n} |C_{\bf n}|^4}}.
\ee
In the limit of a completely localized mode, $R(t)\rightarrow 1$, while for a completely
delocalized state, $R(t)\rightarrow N$, where $N$ is the number of sites in the lattice. For an isotropic square lattice, the modulated state has the form $C_{n,m}=(f_{n} \phi_{n})( f_{m} \phi_{m})$, which means
\begin{eqnarray}
R(t)^{2D}&=&
{( \sum_{n} f_{n}^2 |\phi_{n}(t)|^2  )^2\over{\sum_{n} f_{n}^4 |\phi_{n}|^4}}\times {( \sum_{m} f_{m}^2 |\phi_{m}(t)|^2  )^2\over{\sum_{m} f_{m}^4 |\phi_{m}|^4}}\nonumber\\
&=& (R^{1D}(t))^2.
\end{eqnarray}

\begin{figure}[t]
 \includegraphics[scale=0.45]{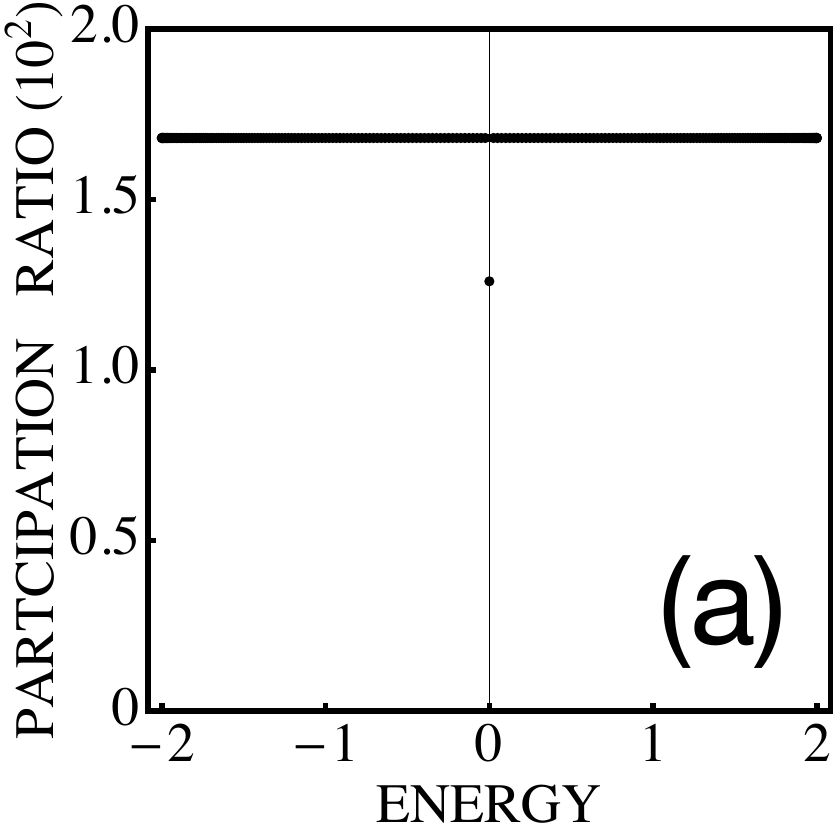}
  \includegraphics[scale=0.2]{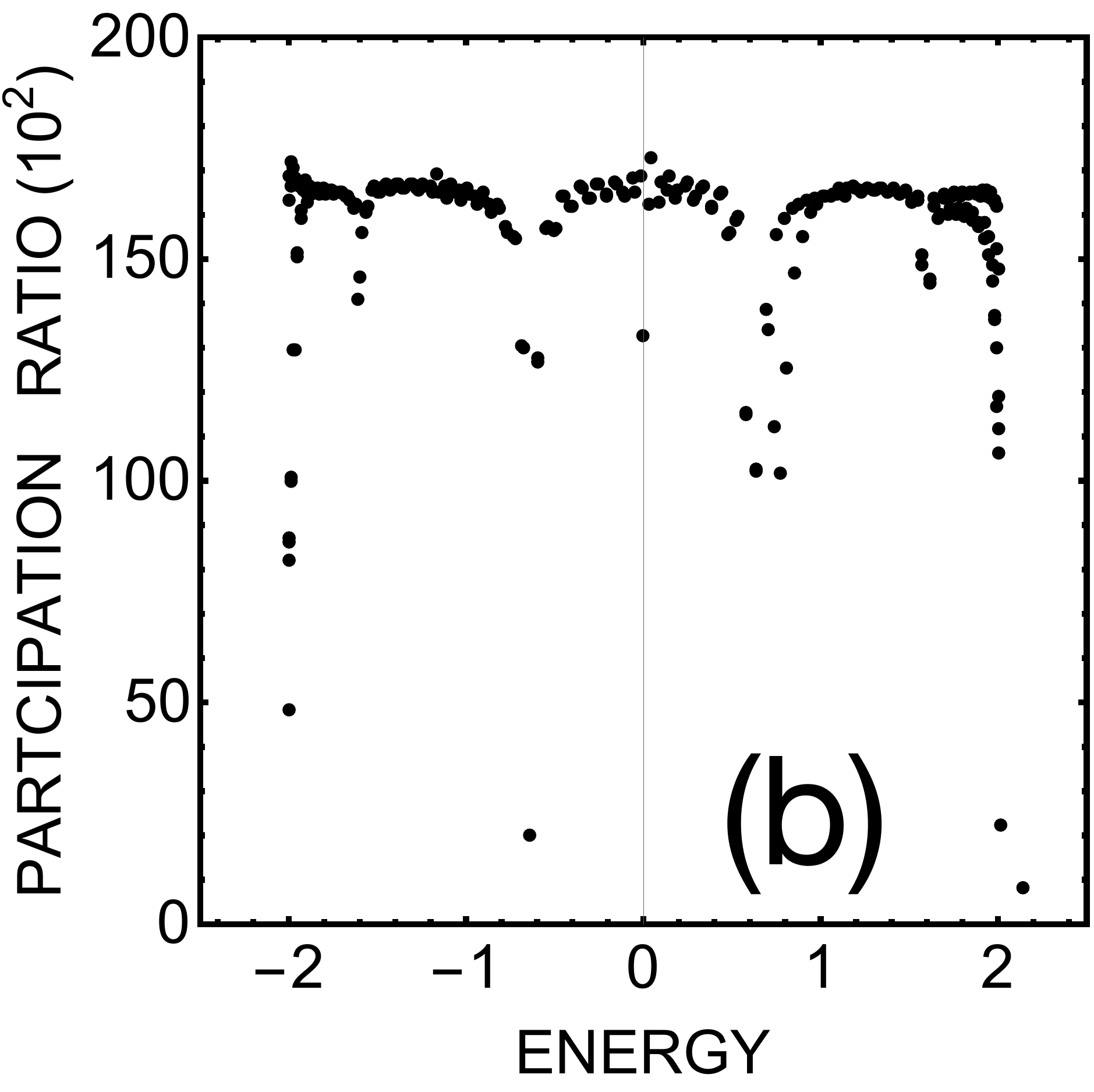}
   \includegraphics[scale=0.145]{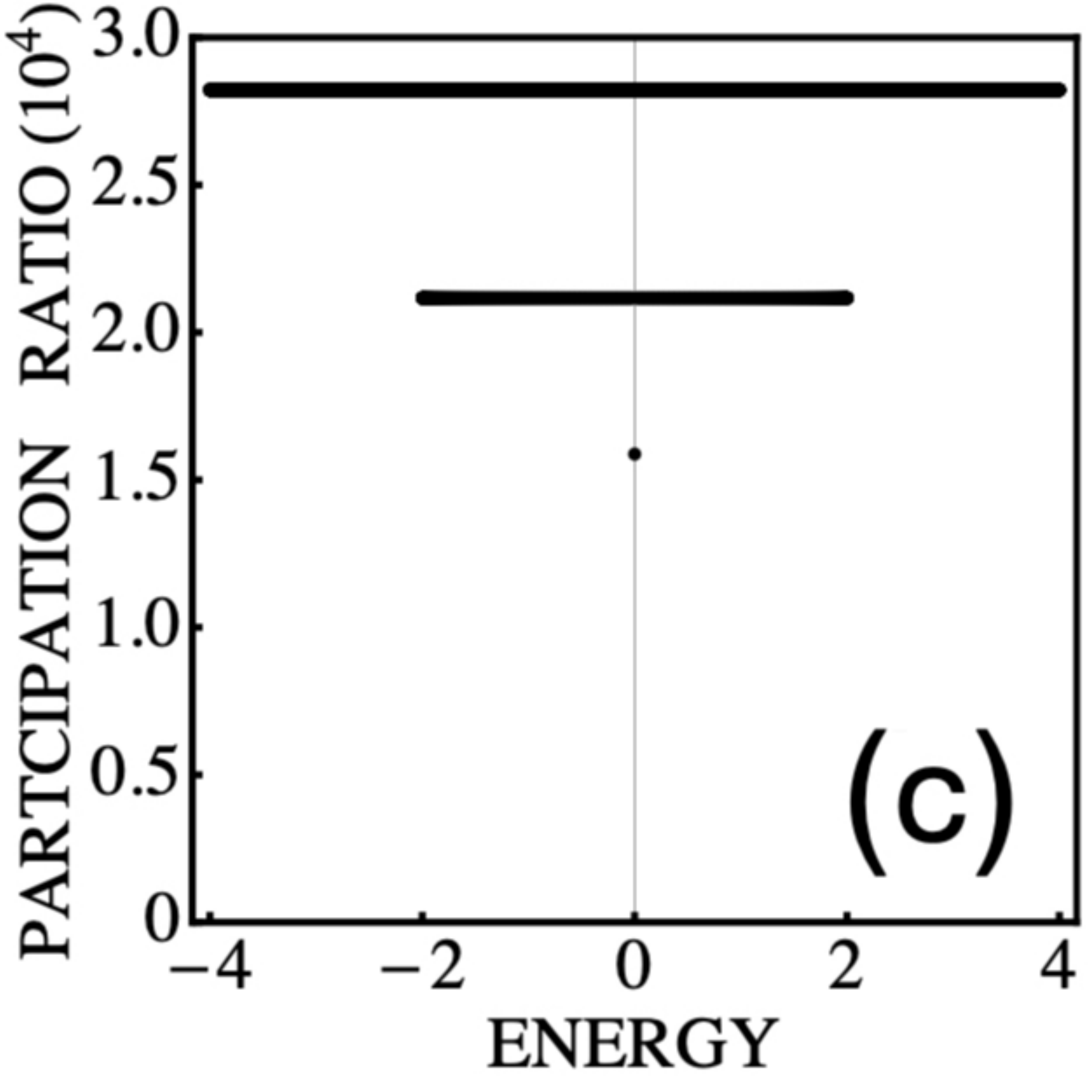}
    \includegraphics[scale=0.14]{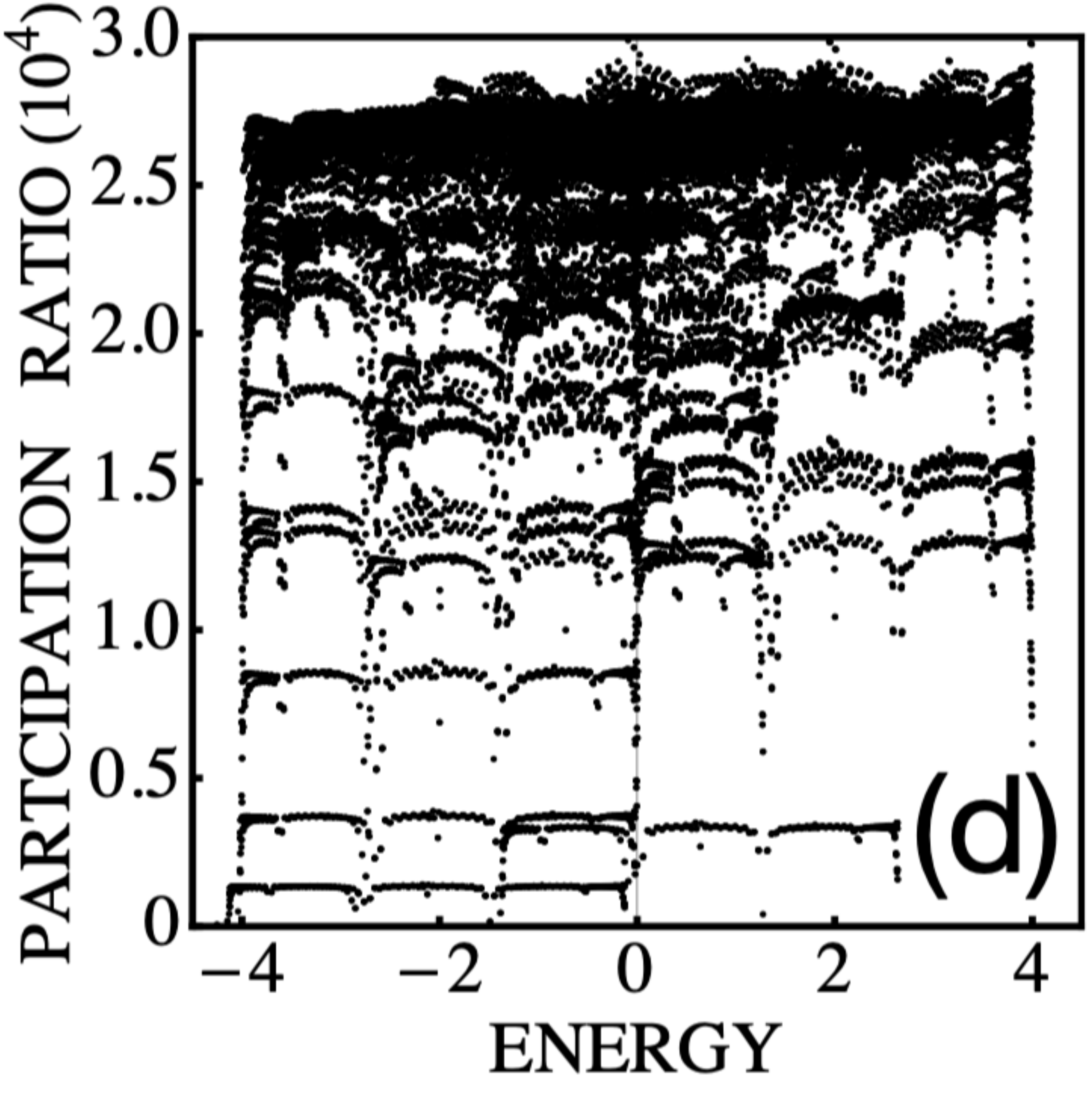}
 \caption{Participation ratio for: (a) 1D lattice with no embedded mode (b) 1D lattice with embedded mode (c) 2D lattice with no embedded mode (d) 2D lattice with embedded mode ($N=251, \lambda=2\times0.63697=1.2739$). }
  \label{fig5}
\end{figure}

For the 1D lattice of $N$ sites, and in the absence of an EM, $\phi_{n}\sim \sin(k n)$ and $\lambda=2 V \cos(k)$, and $R^{1D}$ can be evaluated in closed form as
 \begin{widetext}
 \be
 R(t)^{1D} = {4 N^2\over{3+6 N + \csc(2 k)\sin(2 k(1+2 N))-4 \csc(k)\sin(k(1+2 N))}}
 \ee
 \end{widetext}
 where $k$ is discretized as $k = (\pi/(N+1))\ j$, with $j=1,2,\cdots,N$. 
This can be further reduced to
\be
R(t)^{1D}=\begin{cases}
              (2/3) N\ \ \ \ \mbox{for $k \neq \pi/2$ and $N$ odd or even}\\             
              2 N^2/(4 + 3 N)\ \ \ \ \mbox{for $k=\pi/2$ and $N$ odd}
     \end{cases}
\ee
 Thus, all states of the homogeneous 1D chain are extended, and their participation value scale as $O(N)$ at large $N$. In Figs.\ref{fig5}a,\ref{fig5}b we compare the participation ratio of the 1D lattice with and without an EM. We note that the states at the band edge are affected the most, showing a tendency towards localization. Numerical observations show that most of the modes remain extended but not longer sinusoidal ones. There are also few impurity modes (outside the band) created from the abruptness of the local potential. In Figs. \ref{fig5}c,\ref{fig5}d we show the same comparison but for the 2D case. 
The participation ratio (PR) of the two-dimensional case (Fig\ref{fig5}d), is substantially more complex than its 1D counterpart. Roughly speaking, most of the modes remain extended and contained between the two constant values of Fig.\ref{fig5}c. Below that, one sees a `strata' of one dimensional PR cascading down--from the high to the low-- PR values, ordered like stacked copies of the one-dimensional PR (Fig.\ref{fig5}b). We also notice the presence of several modes with PR smaller than that of our embedded mode. They originate from the 1D impurity states that lie originally outside $[2V,2V]$. The combination of two modes, one inside the band and the other outside, can give rise to a mode inside the 2D 
band $[-4V,4V]$, that is, another embedded mode. By the same token, this combination can also give rise to a mode outside the 2D band ($[-4V,4V]$), i.e., to an impurity mode. To make this discussion more concrete, let us consider a 1D mode inside the band, $\lambda_{1}=2 V-\Delta_{1}$, and another one outside the band, $\lambda_{2}=2 V+\Delta_{2}$. Because of separability, the 2D mode originating from these two 1D modes has $\lambda=4 V-\Delta_{1} + \Delta_{2}$. Thus, an extended 2D mode ($\lambda<4 V$) is generated when $\Delta_{2}<\Delta_{1}$, while a localized mode is created for $\Delta_{2}<\Delta_{1}$. As can be easily surmised, all these 2D modes are a sort of `hybrid modes' being extended along one direction and localized along the perpendicular direction. This picture is valid provided the embedded system is a slightly perturbed version of the original one, and that it stills contains a well-defined band.

To see whether the embedded modes are just impurity modes of a (new) system with several o many gaps, we look at the behavior of the density of states (DOS) $\delta(\lambda) = (1/N^2) \sum_{n,m}\delta(\lambda - \lambda_{n,m})$, where $\delta(x)$ is the Dirac delta function, $\lambda$ is the energy and $\{\lambda_{n,m}\}$ are the eigenenergies of the 2D system containing an EM.  In Fig.\ref{fig6}a we show the DOS of the square lattice with an embedded mode, which is virtually identical to the one belonging to the system without EM (not shown). More importantly, we see no gaps, meaning that the EM is truly embedded in the new band and not an impurity state inside some gap produced by the embedding process.
 
\section{Structural stability}
\begin{figure}[t]
\includegraphics[scale=0.22]{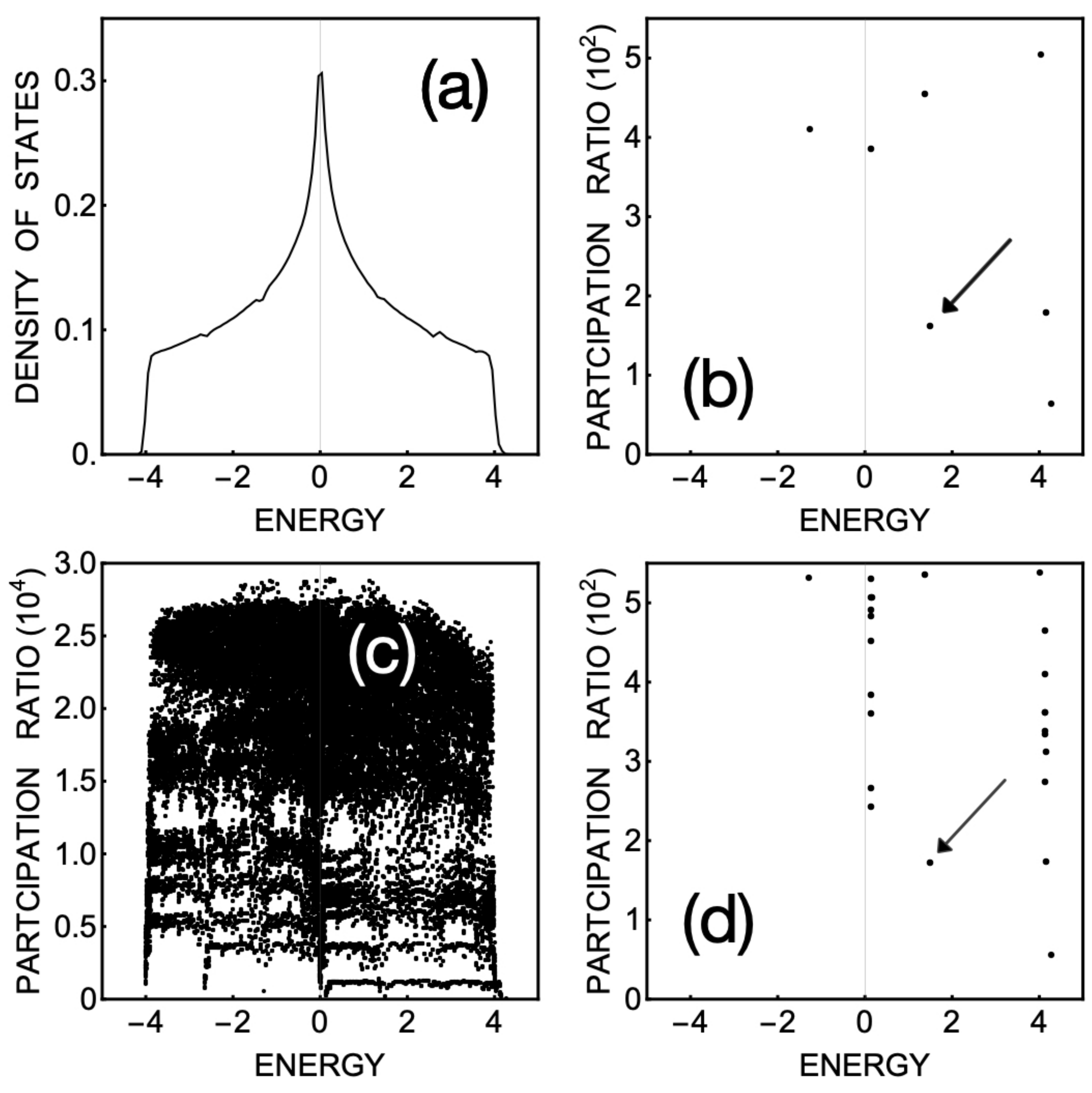}
\caption{(a) Density of states of square lattice with embedded mode.  (b)Participation ratio of embedded square lattice before the random perturbation, at low $R$ values. (c) Participation ratio of embedded square lattice after the random perturbation  (cf. Fig.5d) (d) Same as in (c) but for low $R$ values. The arrows in (b) and (d) mark the position of the EM before and after the random perturbation, respectively ($N=251, \lambda=2\times0.63697=1.2739$). }
  \label{fig6}
\end{figure}

It is interesting to see whether our EM survives the presence of a small random perturbation in the $\{\epsilon_{n,m}\}$. This could correspond to the unavoidable errors in an experimental attempt to fabricate the $\epsilon_{n,m,}$, like an array of optical propagation constants, for instance. Also, we want to see if the structure of the system after the embedding process, is similar to the one before the embedding process. We proceed by changing $\epsilon_{n,m} \rightarrow \epsilon_{n,m} + \xi_{n,m}$, where $\xi_{n,m}$ is random in $[-0.1, 0.1]$, for instance ( this originated from the two onedimensional distributions $[-0.05,0.05]$). Figure \ref{fig6}c shows the participation ratio after the perturbation. In comparison with the case before perturbation (Fig.\ref{fig5}d), we notice that in general, the PR is smeared, with a decreased height meaning a tendency towards localization. The PR shows now a nearly uniform distribution going from high PR to low PR values. The original EM is still there, slightly perturbed in energy, but otherwise robust. In this sense, the EM is stable; however, we also see a number of other localized modes (c.f., Figs \ref{fig6}b and \ref{fig6}d.
that were not there before. As a whole, the PR distribution has definitely changed and in this sense one could say that the square lattice with an embedded mode  is not stable against random perturbations.

\section{Conclusions}
We have examined the creation of an embedded mode (EM) inside the quasi-continuous band of a square lattice, following the original prescription of Wigner and von Neumann. Due to the separability of the problem, it was possible to build  on the previous results for the one-dimensional (1D) chain and thus, construct directly the two-dimensional (2D) envelope function and the 2D potential needed to sustain the EM. The 2D EM is normalizable and decays in space as a stretched exponential, similar to  its 1D counterpart. This is faster than Wigner and von Neumann's case, done for the radial equation of a 3D continuous Schr\'{o}dinger equation. The potential function needed to support the EM decays as a power law at long distances. 
The participation ratio (PR) of all 2D modes was computed and compared to its one-dimensional counterpart. It corresponds to the product of two 1D participation ratios and as such, it shows substantially more complex than in 1D, showing a structure similar to stacked copies of the 1D participation ratio distribution. When the 2D system with a single EM is subjected to a random perturbation, the EM is maintained, but the system suffers a substantial change in its distribution of participation ratio, which gets smeared covering all the range of participation ratio values, and showing an overall reduction in all its values.
This reduced PR is probably an effect of a weak Anderson localization effect due to the complicated shape of the potential that could mimic a disordered system.

For an optics realization of this simple system, the ability to tailor an EM of a given spatial frequency, clearly suggest an application as a spatial filter, since all modes besides the EM are extended and thus possess the ability to propagate through the optical medium; however, if we have an incident wave whose propagation constant $\lambda({\bf k})$ coincides with the $\lambda({\bf k})$ of the EM, it will not propagate. By making a sweep over the chosen $\lambda({\bf k})$ (and its associate potential) it is possible to sweep over the forbidden mode. The most challenging aspect of an eventual optical realization of this system is the fabrication of the optical potential needed to support the EM. This corresponds to a distribution of indices of refraction and its implementation could be laborious given its highly fluctuating nature and its long spatial range.
\vspace{0.5cm}

{\bf Declaration of competing interest}\\
The authors declare that they have no known competing financial interests or personal relationships that could have appeared to influence the work reported in this paper.
\vspace{0.5cm}

\acknowledgments
This work was supported by Fondecyt Grant 1160177.

\end{document}